\documentclass[aps,showpacs,prl,twocolumn]{revtex4}
\usepackage{graphicx}

\begin{document}

\title{Textured Superconducting State in the Heavy Fermion CeRhIn$_5$} 
\author{Tuson Park$^{1, 2}$}
\author{H. Lee$^2$}
\altaffiliation {Present address: Department of Applied Physics and Geballe Laboratory for Advanced Materials, Stanford University, Stanford, California 94305, USA}
\author{I. Martin$^2$, X. Lu$^2$, V. A. Sidorov$^{2,3}$, F. Ronning$^2$, E. D. Bauer$^2$, J. D. Thompson$^2$}
\affiliation{$^1$ Department of Physics, Sungkyunkwan University, Suwon 440-746, Korea\\$^2$ Los Alamos National Laboratory, Los Alamos, New Mexico 87545, USA\\$^3$ Vereshchagin Institute of High Pressure Physics, RAS, 142190 Troitsk, Russia}
\date{\today}

\begin{abstract}
Anisotropic, spatially textured electronic states often emerge when the symmetry of the underlying crystalline structure is lowered~\cite{1}. However, the possibility recently has been raised that novel electronic quantum states with real-space texture could arise in strongly correlated systems even without changing the underlying crystalline structure~\cite{2,3,4,5}. Here we report evidence for such texture in the superconducting quantum fluid that is induced by pressure in the heavy-fermion compound CeRhIn$_5$. When long-range antiferromagnetic order coexists with unconventional superconductivity, there is a significant temperature difference between resistively- and thermodynamically-determined transitions into the superconducting state, but this difference disappears in the absence of magnetism. Anisotropic transport behaviour near the superconducting transition in the coexisting phase signals the emergence of textured superconducting planes that are nucleated preferentially along the $\left\{100\right\}$ planes and that appear without a change in crystal symmetry. We show that CeRhIn$_5$ is not unique in exhibiting a difference between resistive and bulk superconducting transition temperatures, indicating that textured superconductivity may be a general consequence of coexisting orders. 
\end{abstract}
\maketitle

Superconductivity is a macroscopic quantum phenomenon where paired itinerant electrons coherently carry the charge current without any resistance~\cite{6}. Ranging from the discovery of this zero-resistance state to its technological application, electrical resistivity has been a fundamental physical property characteristic of superconductivity (SC). Because a single filamentary path of superconductivity is sufficient to produce zero resistance, this signature of superconductivity can depend sensitively on structural imperfections. Consequently, a broad resistive transition often is taken to indicate poor sample quality~\cite{7}. In strongly correlated electron systems, however, a broad transition can reflect the appearance of a complex state of matter. For example, a broad metal-insulator transition in VO$_2$ arises from correlation-driven nucleation of nanoscale metallic puddles with divergent effective quasi-particle mass~\cite{8,9} and in high-$T_c$ cuprates, the resistive transition can broaden significantly as superconducting, two-dimensional Cu-O planes become electronically decoupled due to a coexisting, electronic phase of stripe order~\cite{10,11}.
 
CeRhIn$_5$ is a strongly correlated antiferromagnet whose Ce-derived 4f-moments order below its Neel temperature $T_N = 3.7$~K at ambient pressure~\cite{12,13}. Application of pressure to CeRhIn$_5$ induces a phase of microscopically coexisting antiferromagnetic order and bulk, unconventional (d-wave) superconductivity for a range of pressures below $P_{c1}$(=1.75 GPa), above which magnetism disappears but superconductivity remains. The width  $\Delta T_c$  of the resistive transition to superconductivity varies strongly with pressure: below $P_{c1}$ , the in-plane resistivity  $\rho_{ab}$ goes to zero at 1.25~K with $\Delta T_c= 0.95$~K at 1.6~GPa, but  $\Delta T_c \leq 0.01$~K above $P_{c1}$ (refs.~\cite{14,15}). The residual resistivity ratio, $\rho_{ab}(300 K)/ \rho_{ab}(0 K)\approx 1000$, is exceptionally large, implying very homogenous crystallinity and that the large  $T_c$ below $P_{c1}$ is not due to poor sample quality but rather is the consequence of a new electronic state. To explore the origin of the broad SC transition, we measured specific heat of CeRhIn$_5$ and its electrical resistivity under pressure for electrical current applied along different crystalline axes. In the coexisting phase, $\rho_c$ sharply goes to zero while $\rho_{ab}$ shows a broad tail. Additional anisotropic transport within the Ce-In plane is consistent with the emergence of textured superconducting planes that are formed preferentially along $\left\{100\right\}$ planes. Disappearence of the anisotropic transport near $T_c$ for $P>P_{c1}$ suggests that the textured SC planes are a consequece of the competing orders of antiferromagnetism and superconductivity.

Single crystals of CeRhIn$_5$ were synthesized by the In-rich flux method\cite{12}. Crystals reported in this article were screened by resistivity and susceptibility measurements that showed no detectable free In wihin the resolution of susceptibility ($\approx0.02$\% volume percent). Conventional four-point contact technique was used to measure the electrical resistivity of CeRhIn$_5$, where samples were polished into bar shapes for different geometry with current being applied along the elongated direction. Specific heat measurements under pressure were performed via ac calorimetry technique~\cite{16}. Quasi-hydrostatic pressure environments were produced by using silicone fluid and glycerol-water mixture (60/40) as pressure transmitting medium for pressure range up to 3~GPa in a clamp-type cell and up to 5.23~GPa in a toroid cell, respectively. Pressures at low temperatures were resistively determined via the suppression of the superconducting transition temperature of Sn or Pb (ref.~\cite{17}).
 
\begin{figure}[tbp]
\centering  \includegraphics[width=7.5cm,clip]{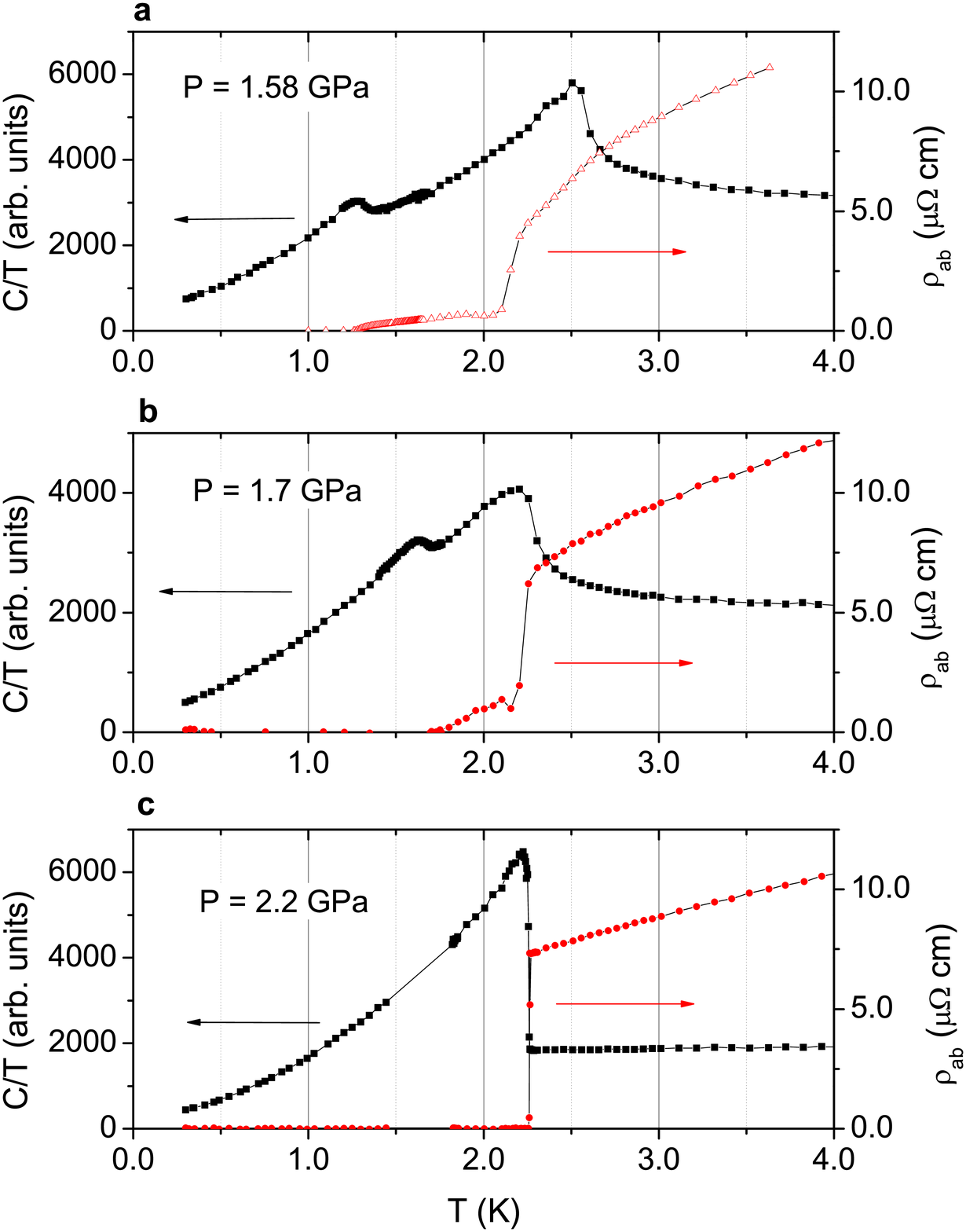}
\caption{(color online) Specific heat divided by temperature ($C/T$) and in-plane electrical resistivity ($\rho_{ab}$) of CeRhIn$_5$ as a function of temperature. $C/T$ and  $\rho_{ab}$ are plotted on the left and right ordinates, respectively. Panels display the data for pressures of 1.58, 1.7, and 2.2~GPa in \textbf{a}, \textbf{b}, and \textbf{c}. At 1.58 and 1.7 GPa, anomalies in $C/T$ at high temperature denote the formation of long-range magnetic order and those at lower temperature are due to bulk superconductivity.  The single anomaly in $C/T$ at 2.2 GPa ($> P_{c1}$) signals the development of superconductivity.}
\label{figure1}
\end{figure}
Figure 1 plots the low-temperature specific heat divided by temperature and in-plane resistivity  $\rho_{ab}$ of CeRhIn$_5$ at pressures of 1.58, 1.70, 2.20 GPa in panels a, b, and c, respectively. At 1.58~GPa, the specific heat shows well-defined anomalies near 2.6 and 1.3~K that correspond to Neel order and to the subsequent bulk superconducting transition $T_c$. The in-plane resistivity, in contrast, reveals three characteristic temperatures of $T_N = 2.6$, $T_{\rho}  = 2.0$, and $T_c = 1.26$~K that are associated with Neel ordering, a dip in the resistivity, and zero resistance, respectively. $T_{\rho}$, clearly a signature of superconductivity, differs significantly from the thermodynamic $T_c$. Instead of going to zero directly, $\rho_{ab}$ shows a long tail of finite resistance down to 1.26~K, the thermodynamic SC transition temperature. The difference, $T_{\rho}- T_c$, shown in Figs.~1b and 1c, decreases with pressure and becomes negligible above $P_{c1}$, indicating that the finite width is related to the presence of the antiferromagnetic order. Scaling of the characteristic temperatures with magnetic field evidences that the same electrons are responsible for the onset, dip, and zero resistance states (not shown).

\begin{figure}[tbp]
\centering  \includegraphics[width=7.5cm,clip]{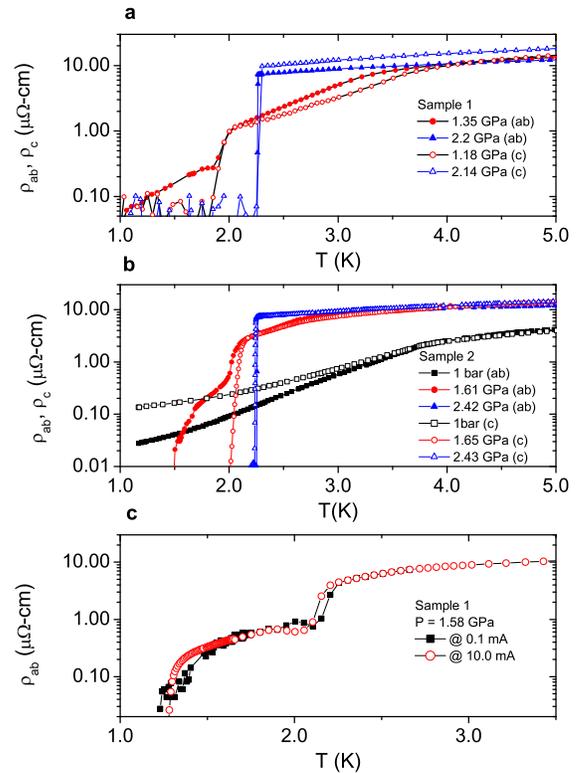}
\caption{(color online) Electrical resistivity of CeRhIn$_5$ for current parallel (open symbols) and perpendicular (solid symbols) to the crystalline c-axis. Data from single crystals of CeRhIn$_5$ from batch numbers 1 and 2 are shown in \textbf{a} and \textbf{b}, respectively. Anisotropy at 1 bar in panel b is associated with antiferromagnetic order. \textbf{c} The in-plane electrical resistivity $\rho_{ab}$ on a semi-logarithmical scale as a function of temperature for applied currents of 0.1 (solid squares) and 10~mA (open circles) at 1.58~GPa.}
\label{figure2}
\end{figure}
Anisotropy between in-plane ($\rho_{ab}$) and out-of-plane resistivity ($\rho_ c$) is plotted as a function of temperature for different pressures in Fig.~2a. Extrapolating these curves to $T=0$ at ambient pressure gives residual values for  $\rho_{ab}$ and $\rho_c$ of $7.9 \pm 0.5$ and $90 \pm7$~n$\Omega \cdot$cm, respectively. If the shoulder-like broad SC transition region were due to sample inhomogeneity or disorder, it would be more prominent for $\rho_c$ because its residual value is much larger than that of  $\rho_{ab}$. Contradicting the expectation, however, $\rho_c$ sharply goes to zero, but $\rho_{ab}$ has an extended tail down to the bulk $T_c$. For $P > P_{c1}$, both $\rho_c$ and $\rho_{ab}$ go to zero simultaneously. Even though the detailed temperature dependence of the shoulder-like feature in the SC region of  $\rho_{ab}$ is sample dependent, the overall characteristic features in the resistivity of CeRhIn$_5$ are reproducible as seen in a comparison of Figs.~2a and 2b. When the electrical current used to measure $\rho_{ab}$ is varied from 0.1 to 10~mA, factor of 10$^6$ change in the input power, the onset of filamentary superconductivity varies slightly due to pair-breaking, but the shoulder-like behaviour persists, confirming the intrinsic nature of the anisotropic resistance-$T_c$ anomaly. 

The disappearance of anisotropy in the resistive transition as well as the coincidence of thermodynamic and resistive transitions above $P_{c1}$ suggest that filamentary superconductivity does not arise from internal structural strains or defects because hydrostatic pressure has no or little effect on these hypothetical extrinsic contributions. Recent neutron scattering measurements have shown, however, that below $P_{c1}$ a new incommensurate magnetic structure Q2 = (1/2, 1/2, 0.391) starts to appear below a characteristic temperature $T^*$ and completely replaces the original one with Q1 = (1/2, 1/2, 0.326) below the bulk superconducting transition temperature $T_c$ (ref.~\cite{18}). The coincidence of $T^*$ with the resistive superconducting onset temperature suggests formation of lamellar superconductivity residing either in the magnetic domain walls or in the nucleated domains of Q2. From the rapid decrease of $\rho_c$ but broad transition in $\rho_{ab}$, we conclude that the domains are preferentially oriented along the c-axis.  Though superconductivity might nucleate first at the domain walls, where both Q1 and Q2 magnetic phases coexist, the observation that the bulk $T_c$ coincides with the transition to the uniform Q2 phase implies that Q2 domains can also support SC.

\begin{figure}[tbp]
\centering  \includegraphics[width=7.5cm,clip]{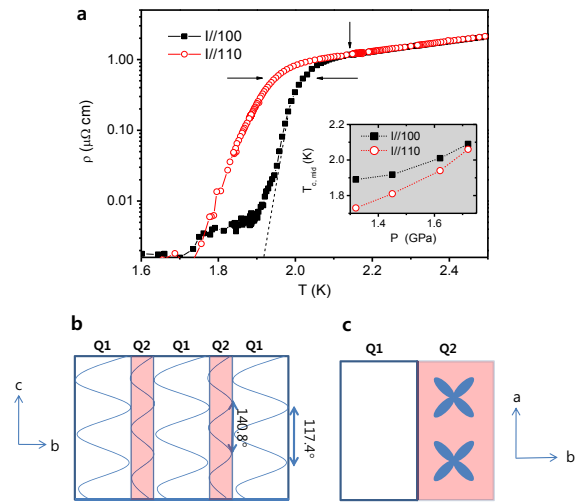}
\caption{(color online) \textbf{a} The in-plane electrical resistivity of CeRhIn$_5$ at 1.45~GPa for current applied along [100] (solid squares) and [110] (open circles). For these experiments, two samples were prepared by cutting one crystal into two pieces, each with different crystallographic orientations. The minimum measurable resistivity of 1.8~n$\Omega \cdot$cm is taken to be zero resistance. Inset: Pressure evolution of $T_{c, mid}$ for the two current directions  [100] and  [110]. $T_{c, mid}$, defined as the temperature at which the resistivity drops to 50 \% of its normal state value, is marked by side arrows and the onset of the SC transition is indicated by a vertical arrow. Dashed lines are guides to eyes. \textbf{b} Schematic illustration of the spatial variation of the amplitude of the textured superconducting state with respect to antiferromagnetic phases with antiferromagnetic ordering wave vectors of Q1=(1/2, 1/2, 0.326) and Q2=(1/2, 1/2, 0.391), respectively. The lines inside the domain describe spatial rotation of the Ce 4f spins. \textbf{c} Illustration of the superconducting d-wave order parameter inside Q2+SC domains, which shows the different pitch angles within the incommensurate Q1 and Q2 phases.}
\label{figure3}
\end{figure}
Figure 3 shows the temperature-dependent resistivity of CeRhIn$_5$ at a representative pressure for electrical current applied along [100] and [110] directions. Even though the onset and zero-resistivity temperatures of the SC transition occur simultaneously for both current directions, the temperature dependence of the transition differs significantly:  [100] initially drops sharply, shows a plateau, and then decreases to the instrumental resolution at 1.7~K, which we take as zero resistance. In contrast,  [110] gradually goes to zero at 1.7~K. The in-plane superconducting anisotropy characterized by the temperature $T_{c,mid}$ and plotted in the inset to Fig.~3b, becomes smaller upon approaching $P_{c1}$ and disappears above $P_{c1}$. The in-plane resistive anisotropy implies breaking of rotational symmetry, with the transition for current flow along [110] being distinct from that of [100] below the resistive SC transition temperature. This is unexpected considering that CeRhIn$_5$ is tetragonal throughout the whole temperature range. A state consistent with these observations is a lamellar structure of Q2 + SC planes embedded in a Q1 matrix. There can be two reasons for anisotropy: either it is caused by non-linear effects in resistivity, or it is due to the presence of domains of size comparable to the sample dimension. Indeed, if the characteristic Q2+SC domain size were much smaller than crystal dimensions, the tetragonal symmetry of CeRhIn$_5$ would not be broken on a macroscopic scale. In this case, the linear resistivity within the ab-plane must be isotropic; in particular, $\rho_{[110]} =\rho_{[100]}$ and any anisotropy could come only from higher order in current terms in resistivity. However, as is shown in Fig.~4, even in the limit of zero applied current, the resistivities in [100] and [110] directions are different, which indicates that the anisotropy is not a result of a non-linear response. Therefore, it must reflect the domain structure with a characteristic length scale comparable to the sample size. If we assume that the magnetic transition from Q1 to Q2 favours formation of domain walls (and hence also domains of Q2+SC) along [100] and [010], then the superconducting sheets will run diagonally to the long axis of the sample oriented along [110] and predominantly parallel to the long axis for sample oriented along [100] (since domain nucleation is biased by the sample surface). Since in the former case there will be no direct path connecting the current source and drain, while in the latter there will, this naturally leads to $\rho_{[100]} < \rho_{[110]}$, as is observed experimentally.

\begin{figure}[tbp]
\centering  \includegraphics[width=7.5cm,clip]{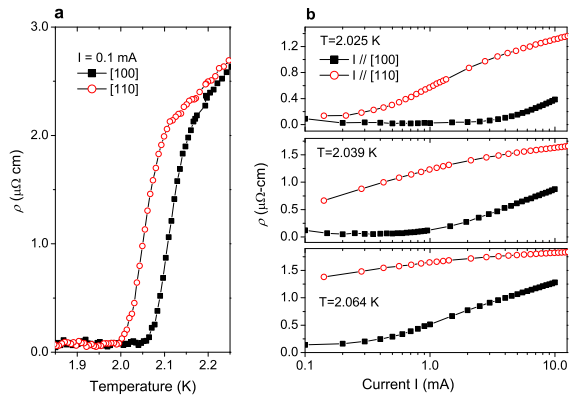}
\caption{(color online) \textbf{a} Electrical resistivity of CeRhIn$_5$ at 1.76~GPa as a function of temperature in the superconducting transition region. Solid and open symbols represent the resistivity for current flow of 0.1~mA along [100] and [110], respectively. \textbf{b} Resistivity of CeRhIn$_5$ at 1.76~GPa as a function of applied current for current flow along [100] (solid symbols) and [110] (open symbols) in the superconducting transition region at temperatures indicated in each panel.}
\label{figure4}
\end{figure}
Striking anisotropy in the superconducting transition of CeRhIn$_5$ suggests the existence of an intrinsic, intermediate region between the bulk and transport $T_c$'s, where the SC state is textured in real space due to a coexisting electronic phase, which in the case of CeRhIn$_5$ is magnetism. Though the study of CeRhIn$_5$ has revealed this texture, there is no reason to suspect that this example is unique. Indeed, it should be a general feature of strongly correlated electron superconductors in which there is a coexisting order. As discussed in SI, similar SC transition pattern is observed when the competing phase is magnetism with incommensurate local moment order (CeRhIn$_5$), strong commensurate order (Cd-doped CeCoIn$_5$), weak incommensurate density wave order (A/S CeCu$_2$Si$_2$), in a superconductor without inversion symmetry (CePt$_3$Si) or in a different class of superconductors (Co-doped BaFe$_2$As$_2$) with density-wave order~\cite{20,21,22,23,24}. Interestingly, as shown by CeIrIn$_5$ and La$_{1-x}$Ba$_x$CuO$_4$ ($x=1/8$), long-range competing magnetic order is not a strict prerequisite for separating bulk and transport transition temperatures~\cite{10, 11, 25}. The observation that an anomalous resistive transition is most prominent in high-quality samples but vanishes in lower-quality specimens contradicts the conventional interpretation that these effects are due to disorder and requires a more fundamental explanation. Experiments, such as those discussed, will reveal details of the coupling and interplay of these orders. 

Work at Los Alamos was performed under the auspices of the U.S. Department of Energy, Office of Science, Division of Materials Science and Engineering and supported in part by the Los Alamos LDRD program. TP acknowledges a support by the National Research Foundation (NRF) grant (2010-0000613 \& 2010-0016560) funded by Korea government (MEST). VAS acknowledges a support from the Russian Foundation for Basic research (grant 09-02-00336).

\newpage
\section{Supplementary Information}
\textbf{We describe supplemental data and analyses that support the textured superconducting state in the heavy fermion compound CeRhIn$_5$. The temperature-pressure phase diagram and scaling behaviour of the upper critical field of CeRhIn$_5$ under pressure are discussed in the Supplement Results. In the Supplement Discussion, we show that CeRhIn$_5$ is not alone in showing a difference between resistive and bulk superconducting transition temperatures, indicating that the textured superconducting state is a common phenomenon in the strongly correlated superconductors with a coexisting competing order.}

\subsection{Supplement Results}
Figure 5 displays the evolution of the antiferromagnetic $(T_N)$ and superconducting $(T_c)$ transition temperatures of CeRhIn$_5$ as a function of pressure at zero magnetic field. When pressure is below $P_{c1}$, the $T_c$ determined from electrical resistivity (open red circles) is higher than $T_c$ from specific heat measurements (solid red circles), while Neel temperatures $T_N$'s determined from both methods are similar~[S1]. For $P>P_{c1}$, where a magnetic state has not been observed, however, superconducting transition temperatures coincide, indicating that the coexistence of magnetism and superconductivity is related to the $T_c$ difference below $P_{c1}$.
\begin{figure}[tbp]
\centering  \includegraphics[width=7.5cm,clip]{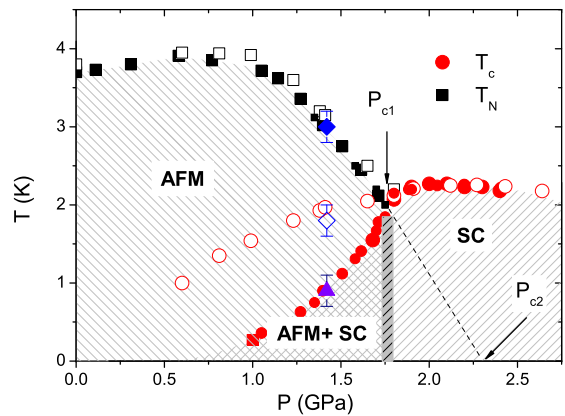}
\caption{(color online) Temperature-pressure phase diagram of CeRhIn$_5$ at zero magnetic field. The superconducting transition temperature $T_c$ (circles) and antiferromagnetic transition temperature $T_N$ (squares) are plotted as a function of pressure. Solid and open symbols are data obtained from the specific heat and the c-axis resistivity measurements ($\rho_c$), respectively [S1]. The onset temperatures of the incommensurate AFM structure Q1=(1/2, 1/2, 0.326) and Q2=(1/2, 1/2, 0.391) of CeRhIn$_5$ at 1.42 GPa, which are obtained from neutron diffraction [S2], are displayed as solid and open diamond symbols. The triangle at 0.9~K is the point where the ICM~Q2 completely replaces Q1.}
\label{figure5}
\end{figure}

Recent neutron scattering measurements show that the incommensurate magnetic ICM~I structure Q1 = (1/2, 1/2, 0.326) develops at $T_N$, while the appearance of the ICM~II phase with Q2 = (1/2, 1/2, 0.391) and the suppression of the ICM~I structure occurs simultaneously at a characteristic temperature $T^*$ (diamond symbols in Fig.~5) (ref.~[S2]). Below the bulk superconducting transition temperature $T_c$, the ICM~II is fully developed, replacing the ICM~I structure. As plotted in Fig.~5, the characteristic temperature $T^*$ (open diamond symbol) coincides with the temperature where the resistance anomaly appears, showing that the magnetic structure Q1 and Q2 coexists between the resistive onset and bulk $T_c$. The fact that bulk superconductivity appears only in the presence of the Q2 structure suggests that the SC electron pairs may be formed in the ICM~II domains or at the ICM I/ICM~II domain walls. Whether the change in magnetic structure arises from superconductivity or the superconductivity is driven by the emergence of ICM~II is not clear at this moment. In the global phase diagram in Fig.~5, however, superconductivity is present over a wide pressure range, including phase space where there is no magnetism. The resistance anomaly $T_c$,  is a smooth function of pressure through the critical pressure $P_{c1}$, indicating that superconductivity is either precipitated on the domain walls between ICM~I and ICM~II, which are free of magnetism, or that ICM~II and superconductivity are only weakly, if at all, competing.

Figure 6a displays the magnetic field dependence of the resistance-$T_c$ anomaly of CeRhIn$_5$ at 1.75~GPa, the critical pressure $P_{c1}$, where antiferromagnetism abruptly disappears at zero magnetic field. The zero-field in-plane resistivity $\rho_{ab}$ starts to drop at 2.3~K, but does not reach zero-resistance immediately. Instead it passes a minimum at 2.2~K, slightly increases with decreasing temperature, then goes to zero at 2.05 K. The appearance of the resistance-$T_c$ anomaly at this critical pressure suggests that there still exists some form of antiferromagnetic correlations among Ce 4f moments. At 5~T where the onset temperature $T_{on}$ of the SC transition is suppressed to 1.85~K, field-induced long-ranged magnetic order clearly is discerned near 2~K and is enhanced with field (marked by an arrow in Fig.~6a). With increasing magnetic field, the $T_c$ anomaly becomes more apparent and there is no complete zero-resistance state at 7~T even though $\rho_{ab}$ starts to drop at 1.5~K. Figure~6b shows the normalized upper critical field $H_{c2}/H_{c2}^{orb}$ as a function of the normalized temperature $T/T_{c0}$, where $T_{c0}$ is the superconducting transition temperature at zero field and $H_{c2}^{orb}$ is the orbital depairing field from the initial slope at $T_{c0}$, i.e., $H_{c2}^{orb} = -0.73T_{c0} (dH_{c2}/dT)T_{c0}$. The orbital critical field is 16.8, 16.1, and 7.8~T for the onset, dip, and zero SC temperature anomalies, respectively. The normalized upper critical fields of the three characteristic transitions collapse on each other, indicating that the same electrons are responsible for the onset, dip, and zero-resistance states.
\begin{figure}[tbp]
\centering  \includegraphics[width=7.5cm,clip]{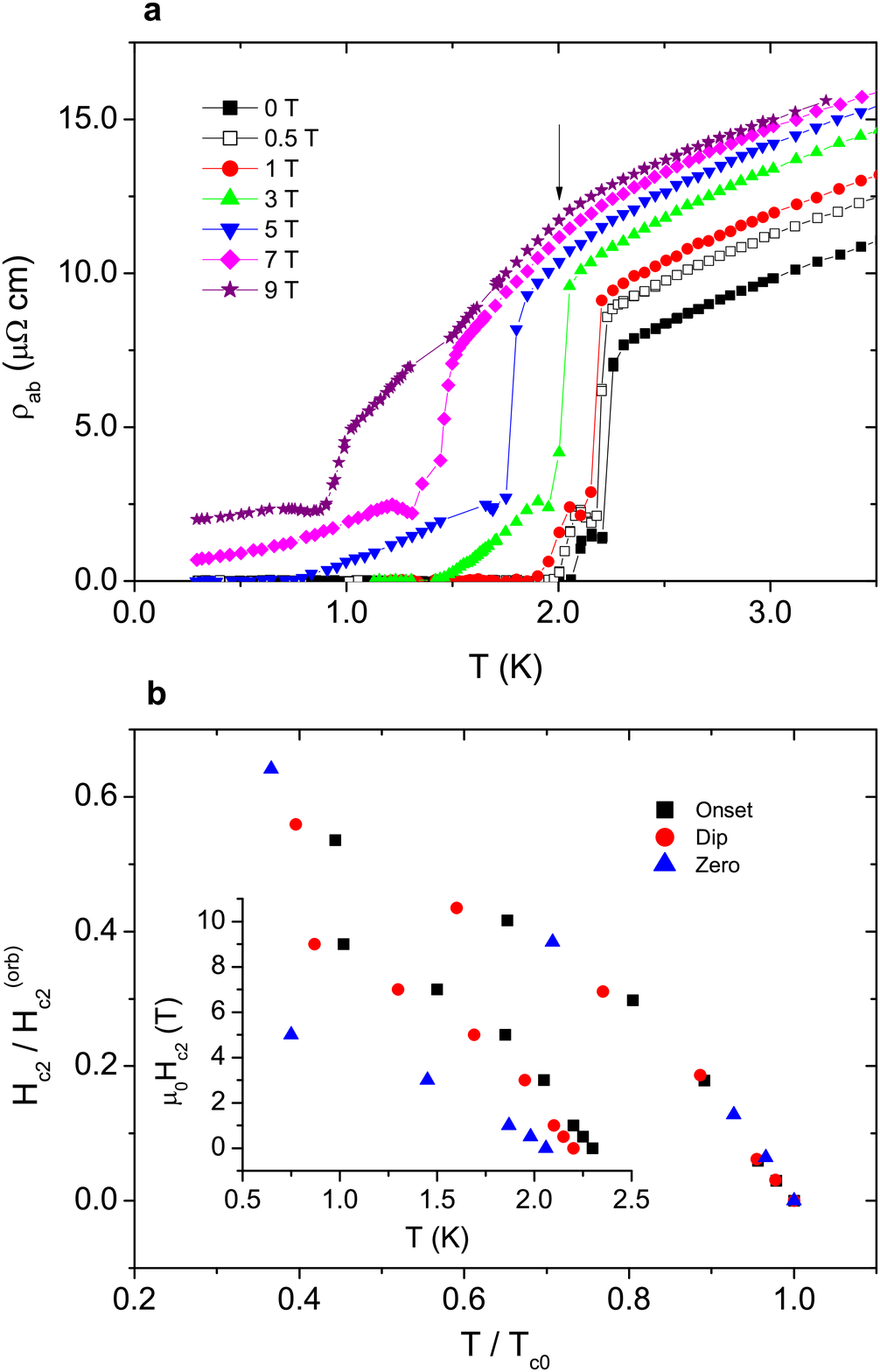}
\caption{(color online) \textbf{a} In-plane electrical resistivity $\rho_{ab}$ of CeRhIn$_5$ at 1.74~GPa for magnetic fields applied within the plane and perpendicular to the electrical current direction: 0 (solid squares), 0.5 (open squares), 1 (circles), 3 (up triangles), 5 (down triangles), 7 (diamonds), and 9~T (stars). The arrow at 9~T marks the field-induced magnetic transition. \textbf{b}  Magnetic field dependence of the three characteristic superconducting transition temperatures of resistance onset (squares), dip (circles), and zero (triangles). In the main figure, the normalized upper critical field $H/H_{c2}^{orb}$ is plotted against the normalized SC transition temperature $T/T_{c0}$, where $H_{c2}^{orb}$ is the estimated orbital depairing field from the initial slope at $T_{c0}$, i.e., 
$H_{c2}^{orb} = -0.73T_{c0} (dH_{c2}/dT)T_{c0}$. The orbital critical field is 16.8, 16.1, and 7.8~T for the onset, dip, and zero SC transitions, respectively. Inset: Magnetic field dependence of the three SC transition temperatures.}
\label{figure6}
\end{figure}

\subsection{Supplement Discussion}
CeRhIn$_5$ is not alone in showing a difference between resistive and bulk superconducting transition temperatures when superconductivity coexists with another order. For example, replacing In by a small amount of Cd in CeCoIn$_5$ induces a phase of coexisting d-wave superconductivity and commensurate antiferromagnetic order with Q=(1/2, 1/2, 1/2). Data  plotted in Fig.~7 at different pressures shows that once antiferromagnetic order disappears at 1.5~GPa, the bulk and resistive $T_c$'s coincide, as in CeRhIn$_5$. 
\begin{figure}[tbp]
\centering  \includegraphics[width=7.5cm,clip]{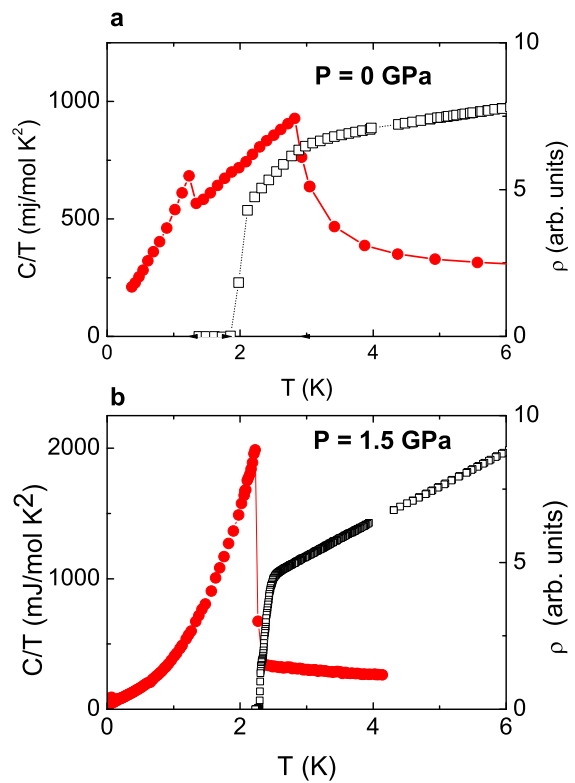}
\caption{(color online) Comparison of the superconducting transition temperatures of 1\% Cd doped CeCo(In$_{1-x}$Cd$_x$)$_5$  that are determined thermodynamically (solid red circles) and resistively (open black squares). Figure~a and~b is at ambient and 1.5~GPa, respectively. Specific heat data is plotted on the left ordinate, while the resistivity is on the right ordinate.}
\label{figure7}
\end{figure}

Unlike the previous examples, CePt$_3$Si crystallizes in a tetragonal structure that lacks a centre of inversion symmetry. In a single crystal of this compound, antiferromagnetic order, with Q = (0, 0, 1/2), forms at 2.2~K and coexists with bulk superconductivity with $T_c = 0.46$~K.  As shown in Fig.~8a, the resistively measured superconducting transition temperature in this crystal appears at atemperature nearly two times higher than the bulk transition [S3].
\begin{figure}[ht!]
\centering 
\begin{minipage} [b] {0.5\linewidth}
\includegraphics[width=6.0cm,clip]{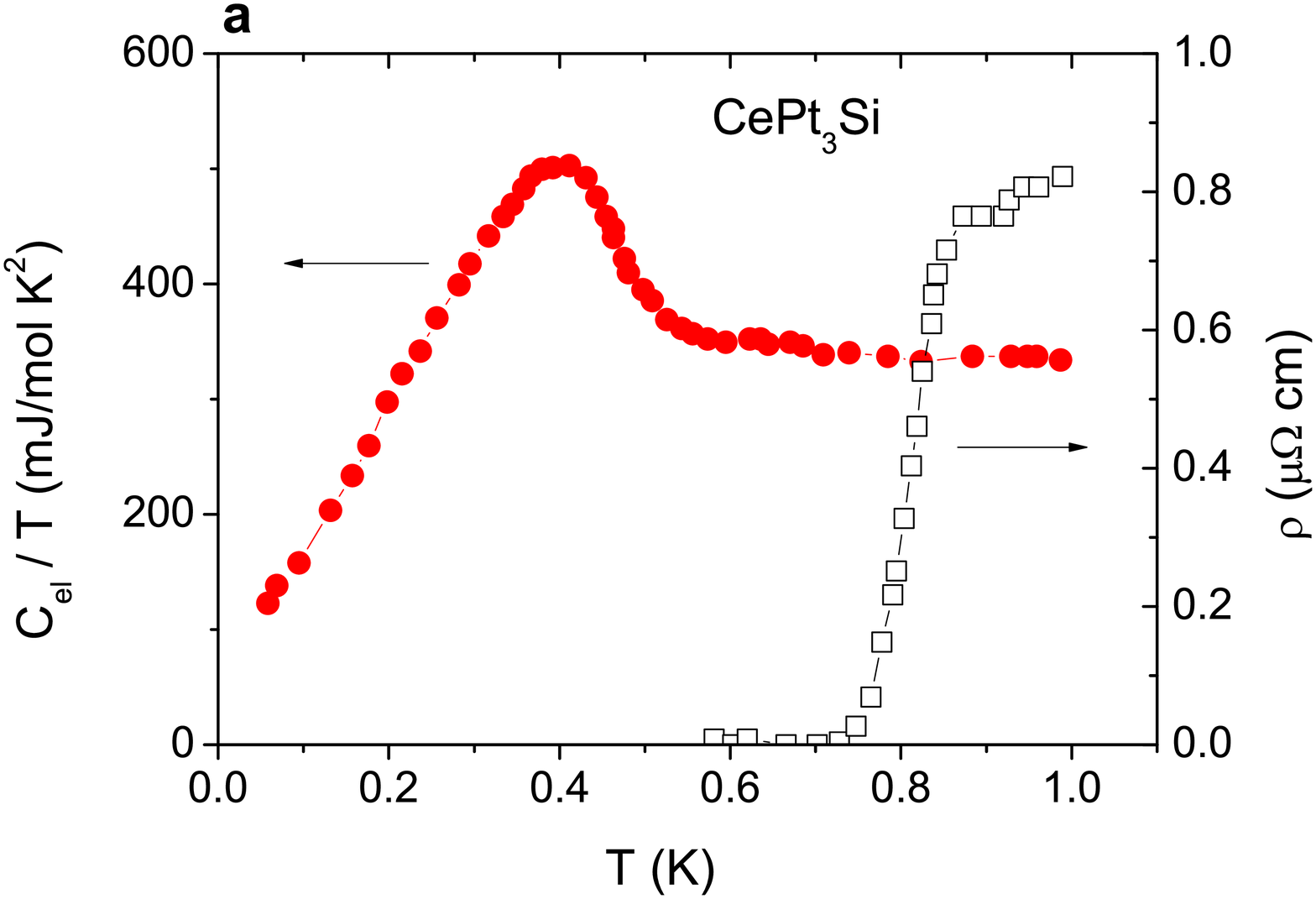}
\end{minipage}
\begin{minipage} [b] {0.5\linewidth}
\includegraphics[width=6.0cm,clip]{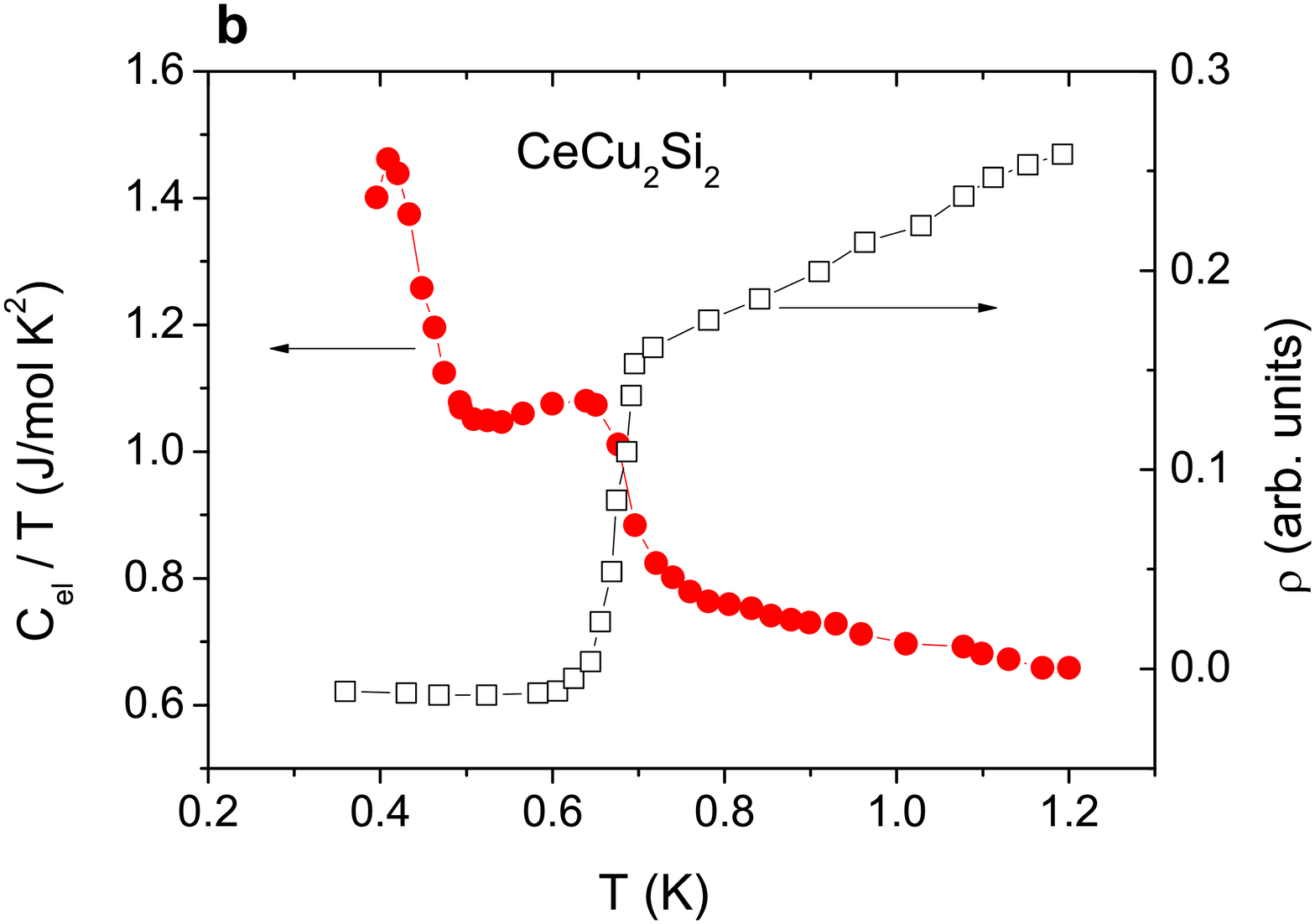}
\end{minipage}
\begin{minipage} [b] {0.5\linewidth}
\includegraphics[width=6.0cm,clip]{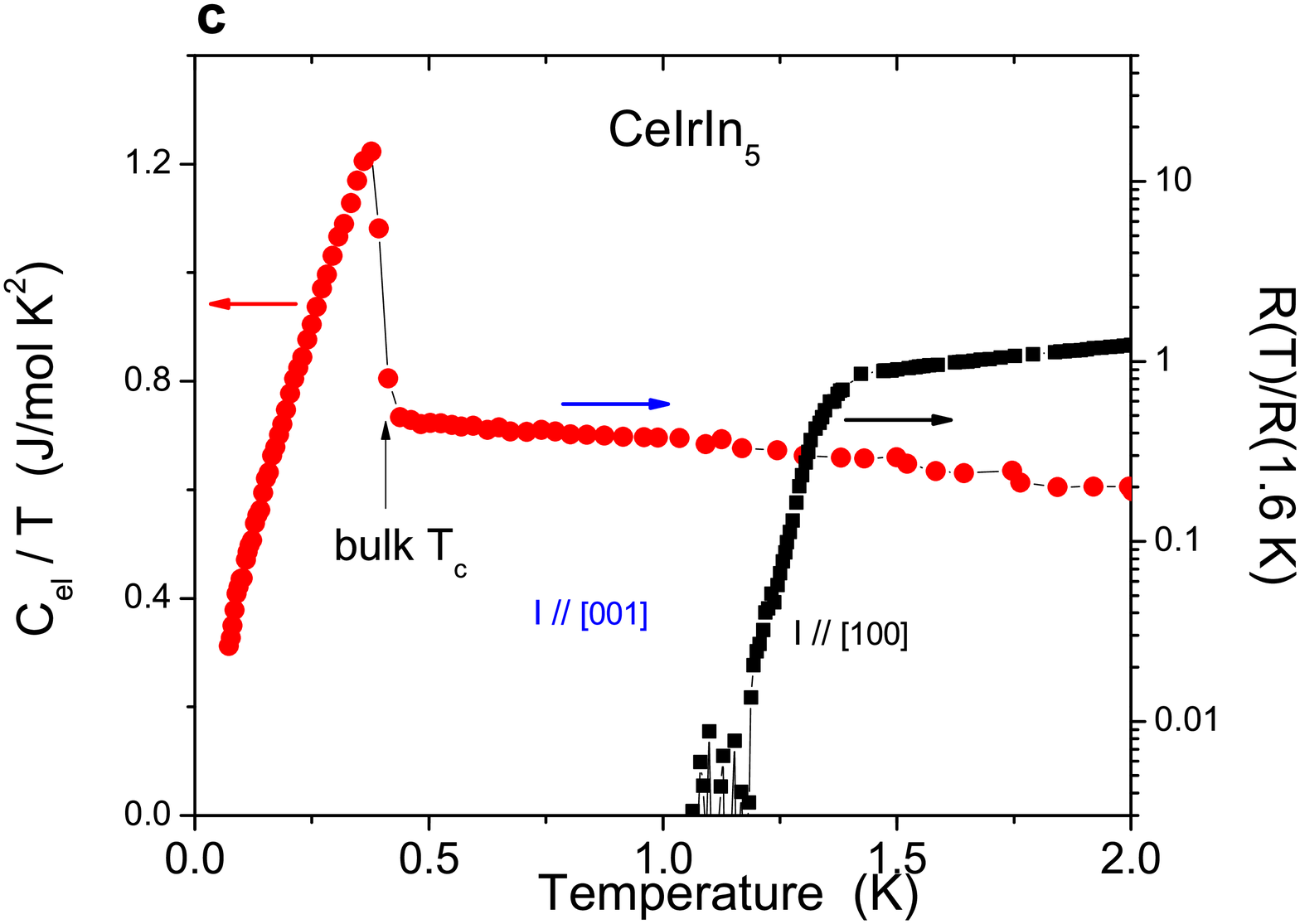}
\end{minipage}
\centering
\caption{(color online) \textbf{a} CePt$_3$Si at ambient pressure. The specific heat (red circles) and resistivity data (black squares) are plotted on the left and right ordinate, respectively. The data are digitized from ref.~[S3]. \textbf{b} Temperature dependence of the electronic specific heat (red circles) and electrical resistivity (black squares) of A/S-type single crystalline CeCu$_2$Si$_2$. Here, A/S describes crystals that show both antiferromagnetic (A) and superconducting (S) transitions. The data are digitized from ref.~[S5]. \textbf{c} Temperature dependence of the electronic specific heat (red circles) and electrical resistance of CeIrIn$_5$ for the current applied along [001] (blue open squares) and [100] (black solid squares). The resistance ratio against that at 1.6~K, $R(T)/R(1.6K)$, is plotted on the right ordinate, while $C_{el}/T$ on the left ordinate.}
\label{figure8}
\end{figure}

A third example is CeCu$_2$Si$_2$ in which incommensurate spin-density wave order develops at $T_N= 0.7$~K (with Q = (0.215, 0.215, 0.530)) and is followed at lower temperature by bulk superconductivity at $T_c=0.46$~K (refs.~[S4], [S5]). In these so-called A/S crystals, bulk superconductivity expels the antiferromagnetic order, but as shown in Fig.~8b, the resistive transition to a superconducting state develops very close to $T_N$. 

In these representative examples, there is always a pronounced difference between resistive and bulk superconducting transition temperatures when antiferromagnetic order is present, irrespective of whether that order is local-moment incommensurate (CeRhIn$_5$), local moment commensurate (Cd-doped CeCoIn$_5$), commensurate but in a structure without inversion symmetry (CePt$_3$Si) or a weak, incommensurate spin-density wave (A/S CeCu$_2$Si$_2$). CeIrIn$_5$, a member of the Ce115 family, also shows a notable difference between resistive and bulk superconducting transition temperatures~[S6]. Though a coexisting order has not been identified yet, Hall effect and magnetoresistance measurements on single crystals find a pseudo-gap-like phase that develops near 2~K in the limit of zero applied field~[S7]. As shown in Fig.~8c, there also is a pronounced anisotropy in the resistive transition to superconductivity. Interestingly, anisotropy in the resistive transition is reversed relative to CeRhIn$_5$, where in CeIrIn$_5$ the transition appears at higher temperatures for current in the tetragonal basal plane. With applied pressure (not shown), the resistive and bulk transitions approach each other and coincide near 3~GPa.  This response is similar to that of CeRhIn$_5$ and Cd-doped CeCoIn$_5$ and strongly suggests that there is a coexisting, but still hidden, order in CeIrIn$_5$.
\begin{figure}[tbp]
\centering  \includegraphics[width=7.5cm,clip]{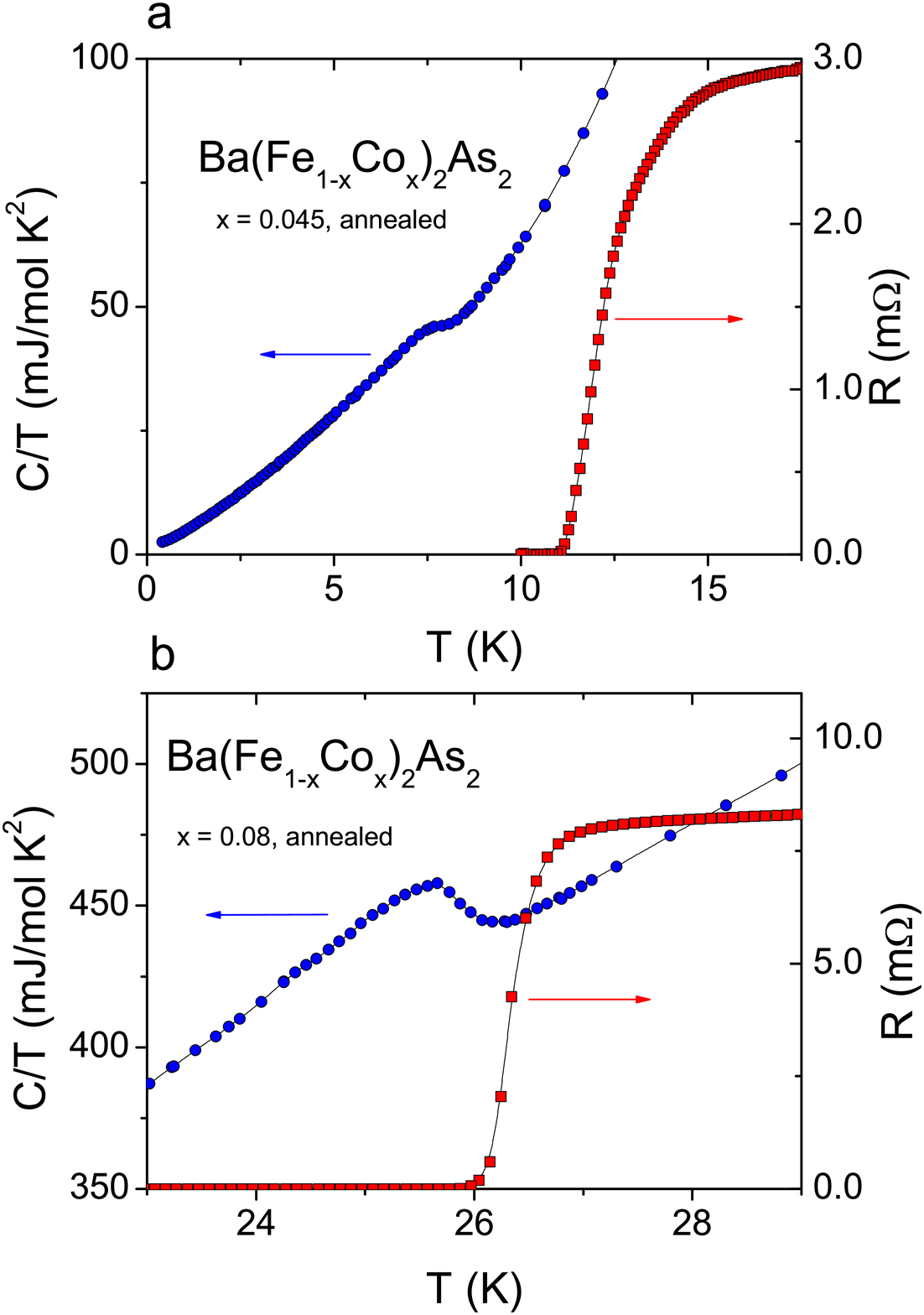}
\caption{(color online) Temperature dependence of the specific heat and electrical resistance of Ba(Fe$_{1-x}$Co$_x$)$_2$As$_2$ with $x=0.045$ and 0.08 in \textbf{a} and \textbf{b}, respectively.}
\label{figure9}
\end{figure}

A conspicuous difference between resistive and bulk SC transition temperatures also is observed in d-electron based superconductors. Figure~9a shows the specific heat and electrical resistance of underdoped (x=0.045) Ba(Fe$_{1-x}$Co$_x$)$_2$As$_2$, with coexisting spin-density wave order~[S8]. In this sample, the zero-resistance state occurs at 11~K but the bulk $T_c$ is 7.8~K. As Co concentration increases close to the optimally doped level ($x=0.08$), where the $T_c$ is highest and antiferromagnetic order is completely suppressed, the difference between the two techniques disappears, i.e., $T_c=26$~K. While further work is needed to determine the intrinsic origin of these observations in the d-electron based superconductors, they support the conclusion that a new textured superconducting state is not unique to Ce-based heavy-fermion materials, but may be a generic feature of strongly correlated electron superconductors where competing orders coexist.

\subsection{Acknowledgments}
Work at Los Alamos was performed under the auspices of the U.S. Department of Energy, Office of Science, Division of Materials Science and Engineering and supported in part by the Los Alamos LDRD program. TP acknowledges a support by the National Research Foundation (NRF) grant (2010-0000613 \& 2010-0016560) funded by Korea government (MEST). VAS acknowledges a support from the Russian Foundation for Basic research (grant 09-02-00336).

\subsection{Supplement References}
[S1] T. Park et al., Proc. Nat. Acad. Sci. \textbf{105}, 6825(2008).

[S2] N. Aso et al., J. Phys. Soc. Jpn. \textbf{78}, 073703 (2009).

[S3] T. Takeuchi et al., J. Phys. Soc. Jpn. \textbf{76}, 014702 (2007)

[S4] F. Steglich et al., Phys. Rev. Lett. \textbf{43}, 1892 (1979).

[S5] E. Lengyel, Thesis (Max-Planck-Institute, Dresden, Germany) (2007).

[S6] C. Petrovic et al., Europhys. Letts. \textbf{53}, 354 (2001).

[S7] S. Nair et al., Phys. Rev. Lett. \textbf{100}, 137003 (2008).

[S8] K. Gofryk et al., (unpublished)


\begin{thebibliography}{10}

\bibitem{1}	E. Fradkin, S. A. Kivelson, M. J. Lawler, J. P. Eisenstein, and A. P. Mackenzie, Annu. Rev. Condens. Matter Phys. \textbf{1}, 153 (2010).
\bibitem{2}	S. A. Kivelson, E. Fradkin, and V. J. Emery, Nature \textbf{393}, 550 (1998).
\bibitem{3}	R. A. Borzi \textit{et al}., Science \textbf{315}, 214 (2007).
\bibitem{4}	Y. Ando, K. Segawa, S. Komiya, and A. N. Lavrov, Phys. Rev. Lett. \textbf{88}, 137005 (2002).
\bibitem{5}	T.-M. Chuang \textit{et al}., Science \textbf{327}, 181 (2010).
\bibitem{6}	J. Bardeen, L. N. Cooper, and J. R. Schrieffer, Phys. Rev. \textbf{108}, 1175 (1957).
\bibitem{7}	A. F. Hebard, and J. M. Vandenberg, Phys. Rev. Lett. \textbf{44}, 50 (1980).
\bibitem{8}	M. M. Qazilbash \textit{et al}., Science \textbf{318}, 1750 (2007).
\bibitem{9}	J. Wei, Z. Wang, W. Chen, and H. Cobden, Nat Nanotechnol. \textbf{4}, 420 (2009).
\bibitem{10}	Q. Li, M. Hucker, G. D. Hu, A. M. Tsvelik, and J. M. Tranquada, Phys. Rev. Lett. \textbf{99}, 067001 (2007).
\bibitem{11}	J. M. Tranquada \textit{et al}., Phys. Rev. B \textbf{78}, 174529 (2008).
\bibitem{12}	H. Hegger \textit{et al}., Phys. Rev. Lett. \textbf{84}, 4986 (2000).
\bibitem{13}	W. Bao \textit{et al}., Phys. Rev. B \textbf{65}, 100505(R) (2002).
\bibitem{14}	T. Park \textit{et al}., J. Phys.: Condens. Matter \textbf{23}, 094218 (2011).
\bibitem{15}	T. Park, M. J. Graf, L. Boulaevski, J. L. Sarrao, and J. D. Thompson, Proc. Nat. Acad. Sci. \textbf{105}, 6825 (2008).
\bibitem{16} Y. Kraftmakher, Phys. Rep. \textbf{356}, 1 (2001).
\bibitem{17} A. Eiling and J. S. Schilling, J. Phys. F: Metal Phys. \textbf{11}, 623 (1981).
\bibitem{18}	N. Aso \textit{et al}., J. Phys. Soc. Jpn. \textbf{78}, 073703 (2009).
\bibitem{20}	L. D. Pham, T. Park, S. Maquilon, J. D. Thompson, and Z. Fisk, Phys. Rev. Lett. \textbf{97}, 056404 (2006).
\bibitem{21}	F. Steglich \textit{et al}., Phys. Rev. Lett. \textbf{43}, 1892 (1979).
\bibitem{22}	E. Lengyel, Thesis (Max-Planck-Institute, Dresden, Germany) (2007).
\bibitem{23}	T. Takeuchi \textit{et al}., J. Phys. Soc. Jpn. \textbf{76}, 014702 (2007).
\bibitem{24}	K. Gofryk \textit{et al}., (unpublished).
\bibitem{25}	C. Petrovic \textit{et al}., Europhys. Letts. \textbf{53}, 354 (2001).

\end{thebibliography}
\end{document}